\documentclass[preprint,12pt]{elsarticle}

\usepackage{graphicx, subfig}

\usepackage{amssymb}
\usepackage{amsthm}
\usepackage{amsfonts,amsmath}

\usepackage{lineno}

\biboptions{comma, square}

\newcommand{\mX}{\mathbf{X}}
\newcommand{\mx}{\mathbf{x}}
\newcommand{\mD}{\mathbf{D}}

\journal{Journal of Theoretical Biology}

\begin{document}

\begin{frontmatter}



\title{Towards Uncertainty Quantification and Inference in the stochastic SIR Epidemic Model}


\author[cimat]{Marcos A. Capistr\'an\corref{corresp}}
\ead{marcos@cimat.mx}
\author[cimat]{J. Andr\'es Christen}
\ead{jac@cimat.mx}
\author[imp]{Jorge X. Velasco-Hern\'andez}
\ead{velascoj@imp.mx}

\cortext[corresp]{Corresponding author}
\address[cimat]{Centro de Investigaci\'on en Matem\'aticas A.C.,Jalisco S/N, Col. Valenciana,
CP: 36240, Guanajuato, Gto, M\'exico}
\address[imp]{Programa en Matem\'aticas Aplicadas y Computaci\'on, Instituto Mexicano del 
Petroleo, M\'exico D.F., M\'exico}

\begin{abstract}
In this paper we introduce a novel method to conduct inference with models defined through
a continuous-time Markov process, and we apply these results to a classical stochastic 
SIR model as a case study. Using the inverse-size expansion of van Kampen we obtain 
approximations for first and second moments for the state variables. These approximate 
moments are in turn matched to the moments of an inputed generic discrete distribution 
aimed at generating an approximate likelihood that is valid both for low count or high 
count data.  We conduct a full Bayesian inference to estimate epidemic parameters using 
informative priors. Excellent estimations and  predictions are obtained both in a synthetic 
data scenario and in two Dengue fever  case studies. 
\end{abstract}


\end{frontmatter}


\section{Introduction}
\label{sec:intro}

%
%

Stochasticity and nonlinearity have a major role in shaping the dynamics of epidemics of infectious 
diseases. It is known that the effects of demographic stochasticity weight more in determining 
the dynamics of epidemics when the number of individuals in the population is low. Consequently, the 
development of mathematical epidemic models that take into account uncertainty and are amenable to describing 
small populations is an active field of research, 
see~\cite{alonso2007stochastic,king2008inapparent,shaw2008fluctuating,nasell2002stochastic,
chen2005stochastic,black2010stochastic}. Among these modeling efforts, models based on Markov processes 
have received a lot of attention. It has been argued that continuous-time discrete-space (CTDS) Markov 
processes whose forward-Kolmogorov equation is known as the chemical master equation (CME) represent the 
stochastic counterpart to systems of ordinary differential equations~\cite{qian2011nonlinear}, and are often 
referred to as birth-and-death processes. The research presented in this paper is based on this modeling 
paradigm. We consider the stochastic SIR epidemic model as a case study. The SIR model and its variants 
are ubiquitous in the study of dynamics of epidemics in populations of plants and 
animals~\cite{bjornstad2002dynamics,gilligan1997analysis}. Seasonality, migration, vaccination,
demographic and environmental stochasticity are among the epidemics features that give rise to variants of 
the SIR model. There is a wealth of qualitative results that have been obtained from the analysis of these 
models, e.g. thresholds for the onset of epidemics, amplification of environmental and demographic noise, 
mechanisms for local extinction and invasion of epidemics. Feedback from the bifurcation analysis of 
deterministic SIR models 
~\cite{kuznetsov1994bifurcation,van2000simple,kribsvelasco2000, alexander2004periodicity} and the analysis of the interplay of
stochasticity and nonlinearity~\cite{rohani2002interplay,dushoff2004dynamical} has provided substantial insight 
regarding epidemic dynamics. A related topic is the quantitative study of the predictive capacity of 
epidemic models.  Recently, parameter estimation of epidemic models, 
from a time series analysis
perspective, and with varying degrees of formality have been
attempted~\cite{finkenstadt2002stochastic,andersson2000stochastic,ramsay2007parameter,munsky2009listening,
archambeau2011approximate}.

Recently, various Bayesian and likelihood based approaches to parameter inference in CME type models
have been attempted.  Some authors have tried to work with a deterministic, continuous or steady state 
alternative model to infer the parameters in the CME \cite{ Rempala:2006, Golightly:2006}, but with 
modest success; specially in the important scenario of low number of counts for the species, when the 
stochastic evolution of the system is in fact the main object of interest \cite{Tian:2007}.
In particular, previous research efforts 
to conduct Bayesian inference with the CME include~\cite{golightly2005bayesian,Golightly:2006},
where the CME is approximated by a diffusion process and~\cite{komorowski2009bayesian}
where the van Kampen expansion is used to derive a diffusion approximation; other similar papers  
include~\cite{ruttor2009approximate,ruttor2009efficient}. However, this approach applies when individual 
counts are large and, in the limit, fluctuations are small and in turn approximated with a Gaussian 
diffusion~\cite[][p. 246]{ruttor2009approximate} which, certainly, will be of limited use if the low
counts for some or all species are an issue, as it is the case with epidemic studies. 

In the chemical master equation scenario simulating data is possible (see Section~\ref{sec.SIR}) and 
the likelihood free Approximate Bayesian Computational (ABC) inference ideas may be used~\cite{Marin2011}.  
The ABC simulates data for fixed 
parameters and rejects trajectories that are not ``close'' (according to an \textit{ad hoc} metric
among an \textit{ad hoc} statistic) to available data.  However, as of yet the ABC lacks of
a firm theoretical support and many crucial details remain unknown, namely, what regularity conditions 
are required (if any) for the ABC to work, how to choose the tolerance and metric and the consequences 
of not using a sufficient statistic~\cite{Marin2011}.  Moreover, when the number of reactions per unit 
of time increases the simulation algorithm may become very inefficient.  \cite{toni2009approximate} 
represents an attempt to use the ABC in this context, for model inference and model selection; however 
a very recent result \cite{robert2011} will severely question the validity of the latter. Other Bayesian 
inference approaches for the CME are~\cite{gillespie2010bayesian, boys:2008} and will be discussed in 
further detail in Section~\ref{sec.BayesInf}.

In this paper we describe a method to conduct parameter estimation in the CME
from partial observation of the state variables. We have used the van Kampen expansion to obtain 
approximate equations of the dynamics of the first two moments of the state variables of the CME
to impute a counting model matching these moments to create a likelihood for Bayesian 
inference. We remark that our results hold for low populations. For the sake of clarity, our examples 
use the simplest epidemic model, e.g. the SIR model without vital dynamics. It is shown that inference 
is possible with the approximate likelihood using synthetic data generated with the Gillespie algorithm 
and the stochastic SIR model. We offer results with data from Dengue Fever onset from two cities central 
Mexico, Acapulco in 2005 and Cuernavaca in 2008. Our contention is that the results presented in this paper 
hold for models formulated as a CME under the hypothesis of spatial homogeneity and thermodynamic equilibrium, 
where only mono-molecular and bimolecular reactions occur.


The paper is organized as follows.  In the next section we explain
the SIR model and the moment approximations that we have used, in Section~\ref{sec.BayesInf} we
explain the details of the Bayesian inference and in Sections~\ref{sec.exa_synth}
and~\ref{sec.exa_real} we present examples with synthetic and real data, respectively.
In Section~\ref{sec.disc} a discussion of the paper is presented.

\section{The Stochastic SIR model and moment approximations}\label{sec.SIR}

Gardiner~\cite{gardiner1985handbook} claims that it is appropriate to describe the dynamics of intrinsically 
discrete systems such as epidemics in terms of jumps. Also, establishes that the chemical master equation 
(CME) offers a complete 
description of such systems since the CME embodies macroscopic deterministic laws of motion about which of the 
stochastic nature of the system generates a random part. However, the chemical 
master equation accounts only for local (demographic) sources of stochasticity. Presumably, a complete model 
of an ecological system such as epidemics should also account for global (environmental) sources of 
stochasticity such as weather. Nevertheless, a general framework to derive stochastic models that takes into 
account both sources of stochasticity is not available. Although it is possible to couple the chemical master
equations with models of extrinsic noise~\cite{alonso2007stochastic,black2010stochastic}, in this paper we 
restrict ourselves to the CME of the classical SIR epidemic model to conduct Bayesian 
inference.

We present briefly the CME as in~\cite{boys:2008}.  For $u$ `species' $\mX = (X_{1}, \ldots ,X_{u})$ and $v$
`reactions' it is assumed that for a small enough time interval $\Delta t$ only one of the following $v$
reactions takes place
$$
R_k : p_{k1}X_1 + \ldots + p_{ku}X_u \stackrel{h_{k}}{\longrightarrow}  q_{k1}X_1 + \ldots + q_{ku}X_u
$$
where $k = 1,2, \ldots , v$ and $p_{kj}$ is the `stoichiometry' of reactant $j$ in reaction $k$ and $q_{kj}$ 
is the `stoichiometry' of product $j$ in reaction $k$, see~\cite{boys:2008} for more details.  This 
means that if reaction $k$ takes place $X_j$ changes by $q_{kj} - p_{kj}$.
Each reaction $R_k$ has an associated reaction rate $h_k( \mX, \theta_k )$ (we assume here that $\theta_k$ 
is a real positive parameter).  Once a reaction has occurred (or at $t=0$) the time to the next reaction has 
an exponential distribution with rate $h_0 ( \mX, \theta_k ) = \sum_{k=1}^v h_k( \mX, \theta_k )$, and the 
$k$th reaction occurs with probability $h_k( \mX, \theta_k )/h_0 ( \mX, \theta_k )$.  This constitutes a pure 
Markov jump process (continuous time) and since only the exponential and a discrete distribution is involved 
it is straightforward to simulate from; this is the so called Gillespie simulation 
method~\cite{gillespie2007stochastic}.
The probability law governing the behavior of this Markov process is the CME.  This model has been applied
in a diversity of fields, e.g. biochemical reactions~\cite{turner2004stochastic}, viral 
infections~\cite{pearson2011stochastic} and ecology~\cite{kot2001elements}.
Although in the following we concentrate on a much simpler setting (the SIR model) we keep this
general model in perspective.  In fact, our inference and prediction procedure may be considered
in this general setting as far as the rates  $h_k( \mX, \theta_k )$ are polynomial, i.e. Kurtz theorem
for the Fokker-Plank approximation holds, see Gardiner~\cite{gardiner1985handbook}.

\subsection{The SIR model}\label{sec.Sirmodel}

As mentioned in Section~\ref{sec:intro}, here we will work on a simple epidemic of the SIR type (susceptible, 
infectious, recovered). The stochastic approach to modeling epidemics has a long history and dates back
to the pioneering work of Kermack and McKendrick \cite{kmk1927, kmk1932, kmk1933}. The main hypothesis of the model is that contact rates occur according to a mass-action law and there is no demographic dynamics, implying that the time scale of the 
model is the length of a single epidemic event. The classical approach has been particularly valuable when 
small population sizes do not support the assumptions of the deterministic ODE models. The reader is referred 
to Daley and Gani \cite{daleygani} for a classical and standard derivation of the basic SIR epidemic model.

Let the random variables $X_{1}$, $X_{2}$ and $X_{3}$ denote respectively the ( ` `species'') number of susceptible, 
infected and recovered individuals in a closed population. The stochastic model is defined by two possible
``reactions'': infection and removal, which occur with ``propensities'' (rates) $h_{1}$ and $h_{2}$ respectively:
\begin{align*}
 R_1: X_{1} + X_{2} &\stackrel{h_{1}}{\longrightarrow} 2X_{2}\qquad\text{infection}\\
 R_2: X_{2} &\stackrel{h_{2}}{\longrightarrow} X_{3}\qquad\text{removal} .
\end{align*}
Here we have $v=3$ species and $u=2$ reactions.  The reaction rates are given by
$h_{1}( X_2, b_0 ) = b_{0}X_{2}/\Omega$ and $h_{2} = b_{1}$, for some positive parameters
$b_0, b_1$ and $\Omega$ to be explained below. 

If we denote by $\mx = (x_{1}, x_{2}, x_{3})$ a realization of the random variables
$\mX = (X_{1}, X_{2}, X_{3})$, and let $P_{(x_{1},x_{2},x_{3})}(t)$ be the probability that the system is in state 
$\mx = (x_{1},x_{2},x_{3})$ at time $t$, then the chemical master equation for this system is given by
\begin{equation}
\label{eq:master_equation}
\begin{split}
\frac{d P_{(x_{1},x_{2},x_{3})}(t)}{dt} &=  b_{0}\frac{(x_{2}-1)}{\Omega}(x_{1}+1)P_{(x_{1}+1,x_{2}-1,x_{3})}(t)\\
        &+b_{1}(x_{2}+1)P_{(x_{1}+1,x_{2}-1,x_{3}-1)}(t)\\
        &-(b_{0}\frac{x_{2}}{\Omega}x_{1} + b_{1}x_{2})P_{(x_{1},x_{2},x_{3})}(t) .
\end{split}
\end{equation}

Let
\begin{align*}
  x_{1} &= \Omega y_{1} + \Omega^{1/2}z_{1}\\
  x_{2} &= \Omega y_{2} + \Omega^{1/2}z_{2}\\
  x_{3} &= \Omega y_{3} + \Omega^{1/2}z_{3},
\end{align*}
where $y_{i}$ y $z_{i}$, $i=1,2,3$ are respectively realizations of the means $Y_{i}$ and the fluctuations 
$Z_{i}$ of the state variables $X_{i}$. Applying the van Kampen's 
inverse size expansion~\cite{gardiner1985handbook} to 
equation~(\ref{eq:master_equation}) leads to (deterministic) 
differential equations for the dynamics of the means
\begin{align}
  \label{eq:sir1}
  \dot{y}_{1} &= -b_{0}\frac{y_{2}}{\Omega}y_{1}\\
  \dot{y}_{2} &= b_{0}\frac{y_{2}}{\Omega}y_{1} - b_{1}y_{2}\\
  \label{eq:sir2}
  \dot{y}_{3} &= b_{1}y_{2}.
\end{align}
Equations~(\ref{eq:sir1})-(\ref{eq:sir2}) represent the macroscopic limit of the CME, 
and coincide with the classical deterministic SIR model. From the van Kampen's expansion we obtain also a 
linear Fokker-Plank partial differential equation governing the dynamics of the fluctuations of the state variables
\begin{equation}
\label{eq:fokker_plank}
\dot{\Psi} = -\sum_{i,j}A_{ij}\frac{\partial}{\partial z_{i}}(z_{j}\Psi)
+\frac{1}{2}\sum_{i,j}B_{ij}\frac{\partial^{2}\Psi}{\partial z_{i} \partial z_{j}},
\end{equation}
where
\begin{displaymath}
 A = \begin{pmatrix}
      -b_{0}y_{2} & -b_{0}y_{1} & 0\\
      -b_{0}y_{2} & -b_{0}y_{1}-b_{1} & 0\\
      0 & b_{1} & 0
     \end{pmatrix},
\end{displaymath}
\begin{displaymath}
 B = \begin{pmatrix}
      b_{0}y_{1}y_{2} & -b_{0}y_{1}y_{2} & 0\\
      -b_{0}y_{1}y_{2} & b_{0}y_{1}y_{2}+b_{1}y_{2} & -b_{1}y_{2}\\
      0 & -b_{1}y_{2} & b_{1}y_{2}
     \end{pmatrix},
\end{displaymath}
and $\Psi = \Psi(z_{1},z_{2},z_{3},t)$ is the probability that the fluctuations are in state 
$(z_{1},z_{2},z_{3})$ at time $t$, see~\cite{van1992stochastic,chen2005stochastic}. From 
equation~(\ref{eq:fokker_plank}) it can be readily established~\cite{chen2005stochastic} that 
the first and second moments of the fluctuations obey the following differential equations
\begin{align}
\label{eq:fluctuations1}
\frac{d\mathbb{E}[z_{k}]}{dt} &= \sum_{k}^{3}A_{ik}\mathbb{E}[z_{k}]\\
\label{eq:fluctuations2}
\frac{d\mathbb{E}[z_{i}z_{j}]}{dt} &= \sum_{k}^{3}A_{ik}\mathbb{E}[z_{k}z_{j}] + 
\sum_{k}^{3}A_{jk}\mathbb{E}[z_{i}z_{k}] + B_{ij},
\end{align}
for $i,j,k=1,2,3$. In the following sections, we shall use 
equations~(\ref{eq:fluctuations1})-(\ref{eq:fluctuations2}) to impute a 
counting model aimed at conducting Bayesian inference with the CME.

\subsection{Synthetic data}

In order to generate synthetic trajectories of the stochastic SIR model in exact accordance with the CME
a simulation method was outlined in Section~\ref{sec.SIR}, namely the Gillespie simulation 
algorithm~\cite{gillespie2007stochastic}.  For a moderate number of reactions, the Gillespie algorithm is feasible.  A sub sampling of
the resulting trajectories will generate data exactly distributed as the CME model.  We will use this procedure
to simulate synthetic data in Section~\ref{sec.exa_synth}. However, the moment approximations
in~(\ref{eq:sir1})-(\ref{eq:fluctuations2}) are used to impute a model for the data and this is used to
perform our inferences.  This in fact avoids 
the testing strategy known as ``inverse crime'', see~\cite{kress1994quasi}, where the same theoretical 
ingredients are used to synthesize and to invert data (infer parameters) in an inverse problem.  We explain this inference
procedure in the next Section.

\section{Bayesian inference and MCMC}\label{sec.BayesInf}

Recently, various Bayesian and likelihood based approaches to parameter inference in CME type models
have been attempted.  Some have tried to work with a deterministic, continuous or steady state alternative model
to infer the parameters in the CME \citep{ Rempala:2006, Golightly:2006}, but with modest success; specially in the most important scenarios
with low number of counts for the species, when the stochastic evolution of the system is in fact    
the main object of interest \citep{Tian:2007, boys:2008}.  Note that, in the artificial extreme case when observations are
available for \textit{all} reactions in \textit{all} species inference is straightforward since the problem becomes one of estimating
the rates in exponential sampling.  If in addition it is assumed that the reaction rates have the form
$h_k( \mX, \theta_k ) = \theta_k g_k( \theta_k )$ even a simple (Gamma) Bayesian conjugate inference
can be conducted~\cite{boys:2008}.

Indeed, data is never complete in real CME applications and the inference problem could be
regarded as one of missing data~\cite{Reinker:2006}.  \cite{boys:2008} takes the latter perspective to develop a
full Bayesian approach.  A consistent handling of missing data is available in the Bayesian paradigm were
a complete data set is simulated from the predictive distribution of missing data given the actual observations and,
in turn, the unknown parameters are simulated form the complete data posterior (the analytic version of this is
totally out of the question in this context).  This creates a two stage sampling
procedure that would generate samples from the correct posterior distribution.  The problem in this context
is simulating complete data sets given observations at fixed time points of typically few (or just one) of the species.
This would mean generalizing the Gillespie simulation algorithm to force all trajectories to pass through the
count observations available for each species.  As opposed to the simplicity of the original Gillespie algorithm, this additional
complication renders the missing data simulation very complex and apparently no direct simulation
method is available.  \cite{boys:2008} attempt two algorithms to approach this missing data simulation
creating a rather complex MCMC.  Even in the case of a very simple two species three reaction CME they obtain
partial success when considering some prior settings and it is indeed not clear how this MCMC will generalize
to other different or more complex CME's~\citep{boys:2008}. 

Since simulating data is possible, the likelihood free ABC inference ideas may be used~\citep{Marin2011}.  
However, as mentioned above, the ABC 
lacks of a firm theoretical support and many crucial details remain unknown, namely, what regularity 
conditions are required (if any) for the ABC to work, how to choose the tolerance and metric, and the 
consequences of not using a sufficient statistic~\citep{Marin2011}.  Moreover, when the number of reactions 
per unit of time increases the Gillespie algorithm may become very inefficient.

Here we take a novel approach to performing Bayesian inference for parameters in the CME, different
in many respects to the previous approaches outlined above.  We use the moment approximations
in (\ref{eq:fluctuations1})-(\ref{eq:fluctuations2}) to impute a counting data model for available 
observations matching those moments, to create an imputed likelihood to perform our inference.  
That is, instead of considering the full Bayesian missing problem approach, we build an approximate 
data model from which inference is much simplified.  Nonetheless, our examples show that inference 
and trajectory prediction in the true CME is possible when using our approximate likelihood. Although 
the Kramer-Moyal moment approximation
would provide similar results as (\ref{eq:fluctuations1})-(\ref{eq:fluctuations2})~\citep[see][chapter 7]{gardiner1985handbook}, 
the van Kampen approximation outlined in Section~\ref{sec.SIR} has a firmer theoretical 
background, and will apply for rate functions common in `chemical' reactions (as those considered here).
Furthermore, the only critical assumption is that the system (the total number of species counts) is
large~\cite{gardiner1985handbook},
which is precisely the case in epidemic models.  That is, individual counts can indeed be low at specific time
points, but the system size $\Omega$ remains large (say $\geq 100$), which in closed population
epidemic models (like the SIR model) remains constant and equal to the population size.
There is also the moment closure approach~\cite{gillespie2007stochastic,gillespie2009} for which a Bayesian 
inference has also been attempted~\cite{gillespie2010bayesian}, but assumes a specific moment structure
and is neither suited to handle low species counts.

\subsection{Approximate Likelihood}\label{sec.approxli}

We assume the difficult case in which only one species is observed.  In most applications of the SIR 
model explained in Section~\ref{sec.SIR} only the number of Infectious individuals at some fixed time 
points are observed (in fact, in some real situation only an approximation for the latter is available, 
see Section~\ref{sec.proxy_data}).  That is, let $t_1, t_2, \ldots , t_n$
be some observation time points and $\mD = ( x_2(t_1), x_2(t_2), \ldots , x_2(t_n))$ the data available.

We define a Generic Discrete Distribution $Gd( m, v)$, that is a combination of the commonly used 
counting data models, namely, the Binomial, the Poisson and the Negative-Binomial, and is explained 
in detail in~\citep{capistran2011generic}. For any mean and variance $\mu, v > 0$ we make a combination of 
these three distributions in the following way
\begin{equation}\label{eqn.gdd}
Gd( x | \mu, v) =
\begin{cases}
C_x^m ~ p^x (1-p)^{m-x};~ & \text{if}~~ \mu > v \\
e^{-v} \frac{v^x}{x!};~ & \text{if}~~ \mu = v \\
C_{x-1}^{x+m-1} ~  p^x (1-p)^m;~ & \text{if}~~ \mu < v \\
\end{cases} 
\end{equation}
where  $x \in \mathbb{N}$, $p = 1 - \min \left\{ \frac{m}{v} , \frac{v}{m} \right\}$, 
$m = \frac{\mu^2}{| \mu - v |}$ and $C_x^m$ 
are the combinations of $m$ items taken in subsets of size $x$.  That is, we use a Binomial if  $\mu > v$, a 
Poisson if $\mu = v$ and a Negative-Binomial if $\mu < v$.  Neither of these distributions can handle 
\textit{any} mean and variance; by combining these distributions we obtain the Generic Discrete class 
$Gd( \mu, v)$ defined for arbitrary mean $\mu$ and variance $v$, $\mu,v  > 0$, and these two moments 
completely define the distribution.
Indeed, it is straightforward to see that if $X \sim Gd( \mu, v )$, $E(X) = \mu$ and $V(X) = v$.  More importantly,
for a fixed mean $\mu$, given both the properties of the Binomial and the Negative-Binomial, we see that
$
\lim_{v \rightarrow \mu} Gd( x | \mu, v ) = e^{-v} \frac{v^x}{x!} .
$  
Therefore we have a continuous evolution of this parametric class, being the Poisson the ``continuous bridge'' between 
the Binomial ($\mu > v$) and Negative Binomial ($\mu < v$). (Note that if $\mu > v$ and  $v \rightarrow \mu$, the 
support will increase to cover all $\mathbb{N}$ since $m \rightarrow \infty$.)
Moreover, if $X \sim Gd( \mu, v )$, $\frac{X - \mu}{\sqrt{v}}$ will tend to a standard Normal distribution if
$\mu \rightarrow \infty$ and $p \rightarrow p_0 \in (0,1)$.  That is, for large $\mu$, large counts, (and for example 
$\mu - 3\sqrt{v} > 0$) $Gd( \mu, v)$ can be approximated with a $N( \mu, v)$. 

From (\ref{eq:fluctuations1})-(\ref{eq:fluctuations2}) we assume that
$\mathbb{E}[X_2(t) | b_0, b_1, \Omega] = m_t$ and $\mathbb{V}[X_2(t) | b_0, b_1, \Omega] = v_t$ are readily
available be numerically solving the referenced system of differential equations.
We assume that
\begin{equation}\label{eqn.impute}
X_2(t_i) \mid b_0, b_1, \Omega  \sim  Gd( m_{t_i} , v_{t_i} )
\end{equation}
(If observations for more species were available, a multivariate generic discrete distribution is also 
explained in~\citet{capistran2011generic} which can be used to define their joint distribution using the cross 
moments in (\ref{eq:fluctuations2})).

Certainly, observations are taken at time points and since we are dealing with a Markov process these are not
independent in general.  The time lagged cross moments $\mathbb{E}[X_2(t)X_2(t+h)]$ would be very useful
in defining the discrete joint distribution for observations, but these moments cannot be calculated in any way similar
to the moments in (\ref{eq:fluctuations1})-(\ref{eq:fluctuations2}). 
Note, however, that the model in (\ref{eqn.impute}) is only an approximation, imputed to match the available moments and,
as approximations go, we may as well take their joint distribution as 
$$
l( \mD ; b_0, b_1, \Omega ) = \prod_{i=1}^{n} Gd( x_2( t_i ) | m_{t_i}, v_{t_i} ) .
$$
Consequently our \textit{approximate} likelihood is defined as the product of individual distributions, as if the
observations were independent.   We present in the next section the corresponding prior and posterior distributions.
The use and validity of this approximate likelihood will only be pondered when we show to recover the true parameter values
in a synthetic data scenario, Section~\ref{sec.exa_synth}, and prove its predictive capability in real data examples,
in Section~\ref{sec.exa_real}.

\subsection{Posterior distribution, MCMC and prediction}

Since $b_0, b_1$ and $\Omega$ are positive parameters, we use a Gamma distribution as a default
and flexible family of priors for these parameters, namely
\begin{equation}\label{eqn.priors}
b_0 \sim Ga( \alpha_0, \beta_0 ), ~~ b_1 \sim Ga( \alpha_1, \beta_1 ) ~~\text{and}~~ \Omega \sim Ga( \alpha_2, \beta_2 ),
\end{equation}
for known hyper-parameters $\alpha_i, \beta_i$.  The log-posterior distribution
is therefore
\begin{displaymath}
\begin{split}
\log f( b_0, b_1, \Omega | \mD ) &= C + \sum_{i=1}^n \log Gd( x_2( t_i ) | m_{t_i}, v_{t_i} )\\
& + \sum_{j=0}^1 (\alpha_j -1)\log(b_i) - \beta_i b_i\\ 
& + (\alpha_2 -1)\log(\Omega) - \beta_2 \Omega ,
\end{split}
\end{displaymath}
for some unknown normalizing constant $C$, $b_0, b_1, \Omega > 0$. 
In some situations the total population size $\Omega$ is known
and the posterior distribution will depend on the parameters $b_0$ and $b_1$ only by fixing the (log) posterior
to $\Omega = N$, for some fixed population size $N$.

The likelihood depends on the parameters $b_0, b_1, \Omega$ through the moments $m_{t_i}, v_{t_i}$.
For any fixed parameter settings, evaluating the likelihood therefore represents numerically solving the differential
equations in (\ref{eq:fluctuations1})-(\ref{eq:fluctuations2}).  Indeed, no analytical alternative is available and therefore
we resort to MCMC simulation techniques~\citep{liu:2001}.

Since no analytical version is available for the posterior distribution, the full conditional distributions cannot be
calculated and usual approaches like the Gibbs sampler are impossible to implement.  
Conventional Metropolis-Hastings MCMC, like a random walk MH~\citep{liu:2001},
are difficult to design and use.  Even a fine tuned MCMC for a particular example may fail radically with the simplest
modification.  We use instead a self adjusted MCMC that automatically tunes itself and has been proved to be of great
use in a variety of situations.  Namely, we use the \textit{t-walk}~\citep{Christen:2010} MCMC sampler (for continuous parameters)
which only requires the log posterior distribution and two initial points for the parameters of interest.  In many examples,
those shown below and many others, the t-walk works quite well in this two or three parameter MCMC simulation problem.

It is very interesting and useful to predict future observations of species; specially in the case of the epidemic example
presented in Section~\ref{sec.exa_real}, where the ability to predict the evolution of the epidemic in the near future may well be
a crucial piece of information for public health decision makers.  In Bayesian statistics the marginal posterior distribution
of observables is calculated to make predictions (i.e.. the posterior predictive distribution).  In this approximate likelihood setting
and once the (t-walk) MCMC sample for our parameters is available, $b_0^{(h)}, b_1^{(h)}, \Omega^{(h)}, h=1, 2, \ldots, H$,
simulating posterior predictive samples for $x_2(t_{n+1})$ is straightforward by simulating from
$Gd( m_{t_{n+1}}^{(h)},  v_{t_{n+1}}^{(h)} )$.  That is, for each of the specific parameter settings $b_0^{(h)}, b_1^{(h)}, \Omega^{(h)}$
the moment approximations are also calculated at the future time $t_{n+1}$ and a (predictive) simulated value
$x_2^{(h)}(t_{n+1})$ is generated by simulating from the Generic Discrete distribution, $Gd( m_{t_{n+1}}^{(h)},  v_{t_{n+1}}^{(h)} )$, explained in Section~\ref{sec.approxli}

\subsection{Handling surveillance epidemic data}\label{sec.proxy_data}

Our aim is to use epidemics surveillance data to make our inferences. According to Diekmann and 
Heesterbeek~\cite{diekmann2000mathematical} prevalence is the number of cases notified up to a given 
time t, and time is measured in some arbitrary scale. On the other hand, the incidence is the expected 
value of the number of new cases per unit time. In order to circumvent the difficulty of using 
surveillance data to estimate the model parameters we have assumed that data is proportional to incidence. 

We assume that the actual number of infectious individuals at reporting times satisfies 
$x_2(t_i) = k r(t_i)$ where $k$ is an unknown constant and $r(t_i)$ are the number of reported cases 
at time $t_i$. That is, the number of infectious people is proportional to the number of reported cases.
Moreover, the initial number of susceptibles $x_1(0) = \Omega - x_2(0)$ is also unknown.  Since $k$ is 
unknown, it is confounded with $\Omega$.   We simply take $\Omega$ as an unknown (now instrumental) 
parameter in our model and take $x_2(t_i) = r(t_i)$.  This pragmatic approach avoids further complicating 
the model and inference and in our synthetic data examples permits reconstructing $b_0$ and $b_1$ even 
when $\Omega$ is also unknown. 

In the examples that we study below cases are reported monthly or weekly and are confirmed through an 
official detection algorithm approved by the Federal Health Secretariat of Mexico. In this respect the 
Cuernavaca data is probably the one better monitored since information on each case is more complete 
(age, sex, address, reference numbers, dates of admission to healthcare, etc are provided). Nonetheless, 
both examples are only meant to test our methods and therefore show probably two typical cases of the 
extremes of data quality and availability. In a forthcoming paper we will be more concerned with the 
actual population dynamics of the disease where a more detailed data selection and treatment is necessary.

\section{Example: Synthetic Data}\label{sec.exa_synth}

\begin{figure}
\centering
\subfloat[]{\includegraphics[height=8cm, width=14cm]{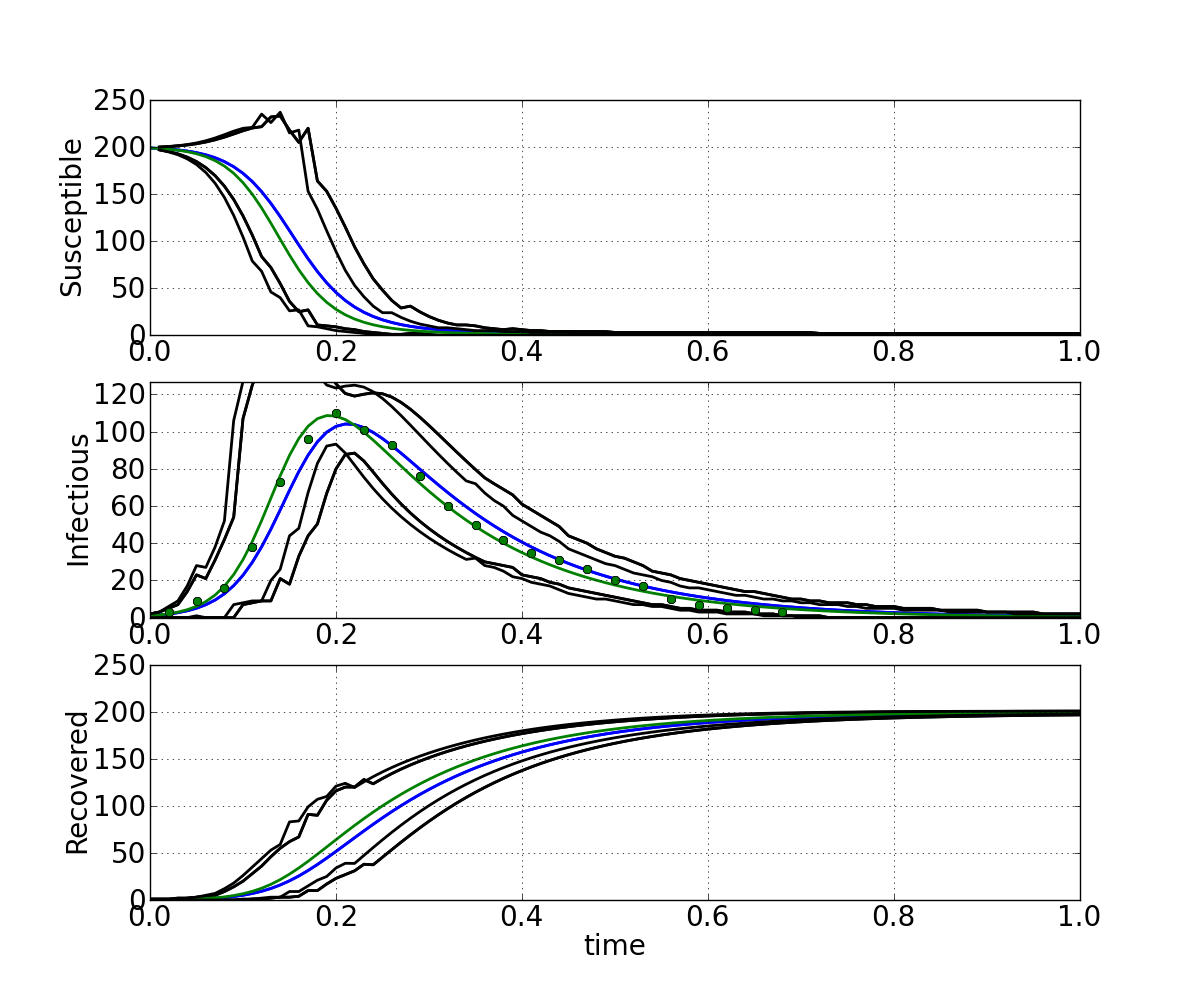}} \\
\subfloat[]{\includegraphics[height=4cm, width=5cm]{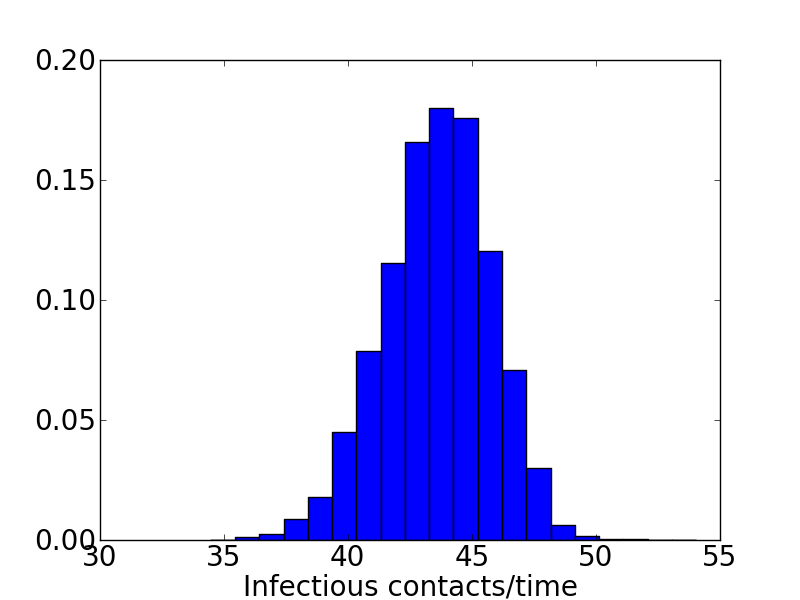}}
\subfloat[]{\includegraphics[height=4cm, width=5cm]{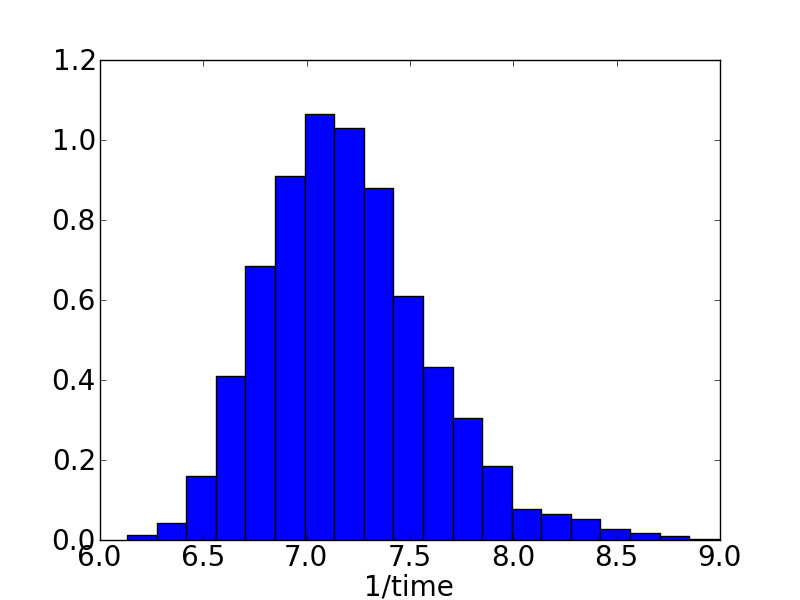}}
\subfloat[]{\includegraphics[height=4cm, width=5cm]{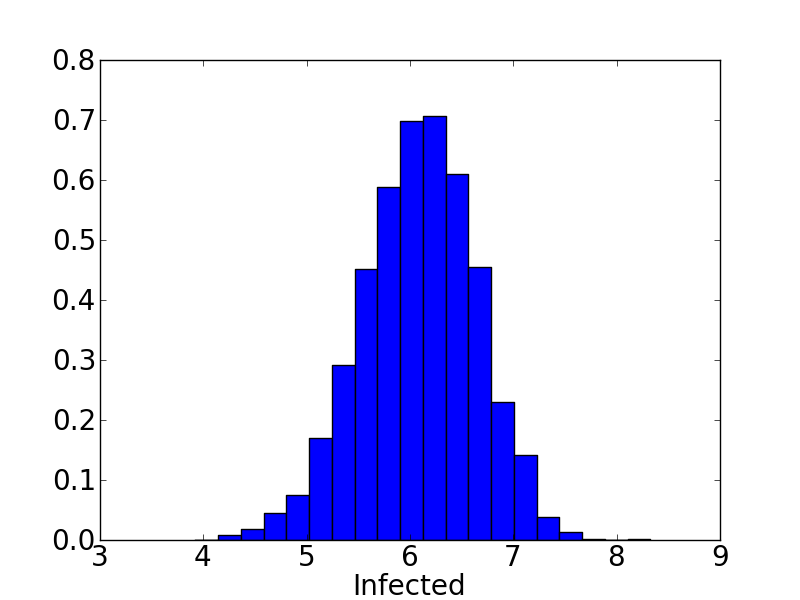}}
\caption{\label{fig.synth1} Synthetic data with $b_0=40, b_1=7$ and $N=200$.  (a) True model (blue) and
MAP estimated model (green) and 5\% and 95\% quantile bands for the imputed $Gd$ distribution at any time.
Posterior distributions for $b_0$  (b), $b_1$ (c) and $R_0$ (d).}
\end{figure}

\begin{figure}
\centering
\subfloat[]{\includegraphics[height=8cm, width=14cm]{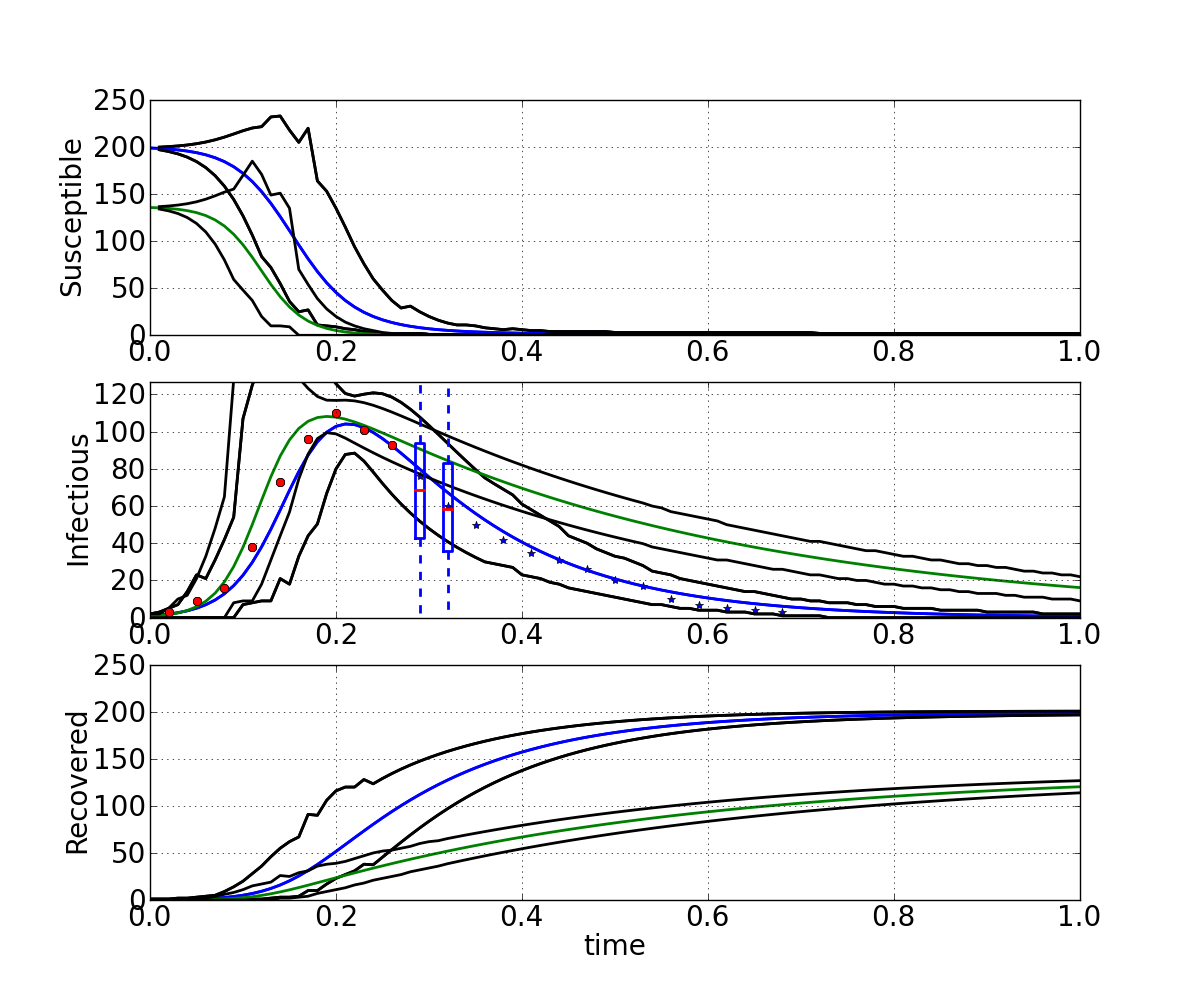}} \\
\subfloat[]{\includegraphics[height=4cm, width=5cm]{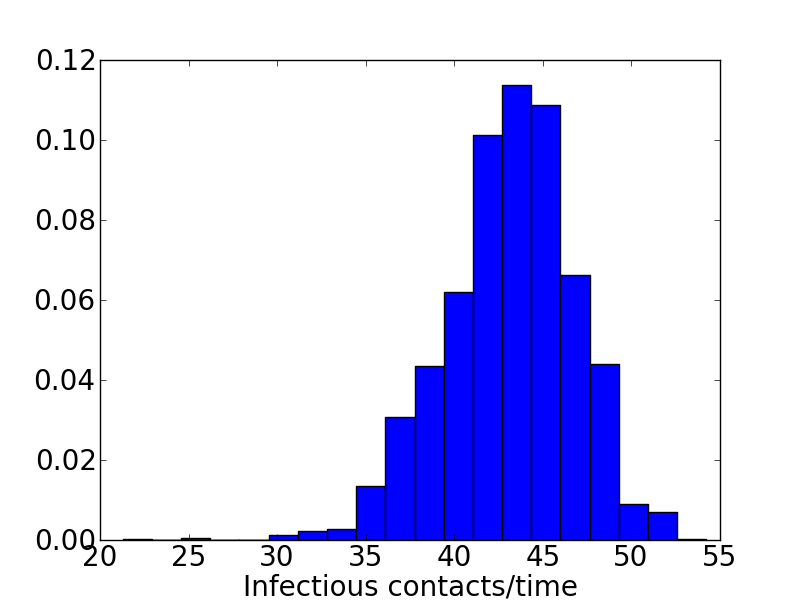}}
\subfloat[]{\includegraphics[height=4cm, width=5cm]{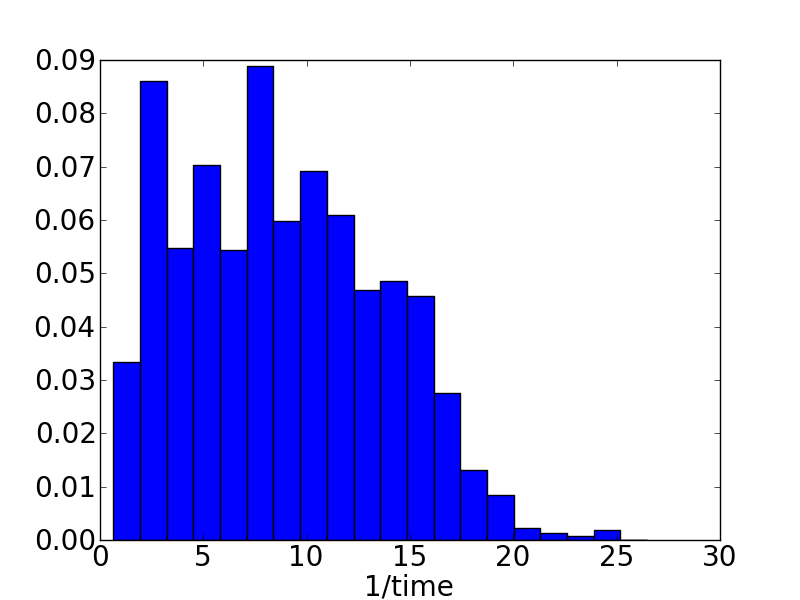}}
\subfloat[]{\includegraphics[height=4cm, width=5cm]{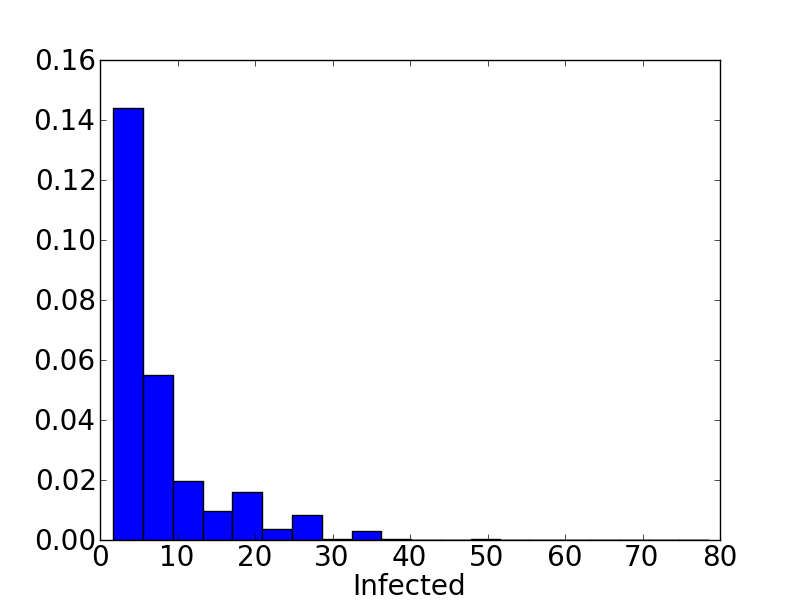}}
\caption{\label{fig.synth2} Synthetic data with $b_0=40, b_1=7$ and $N=200$, 14 observations
trimmed away from the end.  (a) True model (blue) and
MAP estimated model (green), 5\% and 95\% quantile bands and box-plots for
the posterior predictive distribution at two future points (actual data in *).
Posterior distributions for $b_0$  (b), $b_1$ (c) and $R_0$ (d).}
\end{figure}

In this section we offer synthetic examples aiming at showing the accuracy and predictive features of the 
inference methods developed in the previous sections. For the sake of conciseness, we restrict ourselves 
to two contrasting cases. First, we set $\Omega = 200$ and use many data points to solve the inference problem, 
see Figure~\ref{fig.synth1}. In the second example $\Omega$ is regarded as unknown (also a parameter) and many 
data points are trimmed out of the analysis. Furthermore, some data points are predicted, see 
Figure~\ref{fig.synth2}. In both examples synthetic data was created with the Gillespie algorithm mentioned 
above and fixed parameter values to $b_0=40$ and $b_1=7$.

By using the formalism of the CME we are allowing for stochastic fluctuations in the state variables that, 
in the case of the SIR model, refer to demographic stochasticity due to individual differences in contact 
rates ($b_0$) and time of infectiousness ($b_1$).  In Figure~\ref{fig.synth1}(d) we present the posterior 
distribution for $R_0$ which, therefore, approximates the distribution of $R_0$ among individuals in the 
population.  The Maximum \textit{a posteriori} (MAP, the posterior mode) estimator for $R_0$ is 6.0, with 
$( 4.8, 7.0)$ as Highest Posterior Density 95\% probability interval~\citep[HPD, an interval accumulating 
0.95 probability with minimum length, the Bayesian equivalent of `credible' intervals, see][]{bernardo1994}.
Model predictions follow the true epidemic behavior with reasonable accuracy (see Figure \ref{fig.synth1}(a)). 
In this experiment the full data set was used. 

To test for robustness we performed a new run but trimming away the last 14 weeks of the epidemic and considered the total population size $\Omega$, as unknown. Essentially we are only using the information necessary for the estimation of $R_0$. Model predictions are presented in Figure \ref{fig.synth2}(a). The capacity of the model to predict the epidemic tail is remarkable although, as expected, the uncertainty on the estimation of $R_0$ increases relative to the previous case, moreover,
its posterior distribution is quite skewed, see Figure~\ref{fig.synth2}(d).
However, although in this case the MAP estimator for $R_0$ is 2.9, its posterior mean is 6.3, quite similar to the estimator of $R_0$
in the complete data, known $\Omega$, previous example.  The 95\% HPD probability interval is $( 2.9, 16.2)$. Note that for this skew distribution a single estimator is a very bias representation of the posterior distribution.
Unlike all other posterior densities presented here
the MAP and posterior mean differ in this case.  The posterior distribution itself should always be used for 
a better interpretation in this Bayesian analysis~\citep[a point that went unnoticed in][]{coelho2011}.

\section{Example: Dengue Fever Outbreaks}\label{sec.exa_real}

There are numerous studies on the estimation of the basic reproductive
number for Dengue. Many of these works look at the local initiation
of the outbreak to estimate directly the basic reproductive number
(e.g., Chowell, Diaz-Due\~nas, Miller et al,  \cite{chowell2007}; Mendes-Luz, Torres-Codeco,
Massad et al,  \cite{mendes2003}) or use a phenomenological approach to approximate
its upper bound  (Hsieh and Chen,
 \cite{HsiehChen2009}; Hsieh and Ma,  \cite{hsiehma2009}).  Here
we are able
to estimate the basic reproduction number and the fate of the epidemic for the Dengue outbreaks in Acapulco, Mexico, in
2005 and Cuernavaca, Morelos in 2008 using a full differential equations model as
explained earlier.

In Table~\ref{tab:R0s} we list some estimates
for the reproductive number for Dengue with their corresponding confidence
intervals (when available) and for different cities. The reader is referred
to the cited reference for further details. We have omitted those references
whose estimations show very large variability.

\begin{table}
\caption{\label{tab:R0s} Reported data on the basic reproduction number for Dengue from several 
sources. Adapted from Hsieh and Chen \cite{HsiehChen2009}}
\begin{tabular}{|c|c|c|}
\hline 
Source & $R_{0}$ or range & Confidence interval 95\%\tabularnewline
\hline
\hline 
Hsieh and Ma \cite{hsiehma2009} & $2.23$ & $(1.47,3.00)$\tabularnewline
\hline 
Hsieh and Chen \cite{HsiehChen2009} & $3.93-4.67$ & \tabularnewline
\hline 
Koopman et al \cite{koopman1991} & $1.33-2.40$ & \tabularnewline
\hline 
Marques el al \cite{marques1994} & $1.6-2.4$ & \tabularnewline
\hline 
Khoa et al \cite{khoa2005} & $1.25-1.75$ & \tabularnewline
\hline 
Chowell et al \cite{chowell2007} & $3.09$ & $(2.34-3.84)$\tabularnewline
\hline 
Chowell et al  \cite{chowell2007} & $2.0$ & (1.75-2.23)\tabularnewline
\hline 
Chowell et al  \cite{chowell2008} & $1.76$ & \tabularnewline
\hline 
Massad et al  \cite{massad2008} & $1.9$ & \tabularnewline
\hline
\end{tabular}
\end{table}

In the 2005 and 2008 epidemic outbreaks recorded for the cities of Acapulco, state of Guerrero,
and Cuernavaca, state of Morelos, Mexico, 
respectively only one viral strain was identified (no co-circulation of various
serotypes). In 2005 DEN-1 and DEN-2 in Guerrero was isolated (Carrillo-Valenzo, Danis-Lozano,
Velasco-Hern\'andez et al, \cite{carrillo2010} and data from Direcci\'on General de Epidemiolog\'{\i}a) and in 2008 in Morelos it was DEN-2 (Direcci\'on General de Epidemiolog\'{\i}a)

By using the SIR model we are neglecting the role of the population dynamics of the mosquito 
during the epidemic outbreak and are, therefore, incorporating its effects as part of the 
constant contact rate $b_1$.  
Our rationale behind this assumption is the following: during the epidemic
period the density of mosquito increases substantially. Smith, Dushoff
and McKenzie  \cite{smith2004} have shown that during an epidemic outbreak the
human biting rate is highest shortly after mosquito density peaks
and that the proportion of infected mosquitoes peaks when mosquito
population density is declining. On the other hand Sanchez, Vanlerberghe,
Alfonso et al  \cite{sanchez2006} show data that confirms that mosquito abundance
is strongly correlated with an epidemic. There have been reports for Rio de
Janeiro, however
(Honorio, Nogueira, Codeco et al, \cite{honorio2009}), that have shown that the time of highest case 
incidence is not associated with
a local increase of vector abundance, observation that can be caused
by the movement patterns of the population. In this work we are looking
at a geographical scale that comprises large metropolitan areas where
the environment for transmission, during an epidemic, can be thought
of as saturated with mosquito although at a local level (that of city
district, block or house), it might not be the case. A high mosquito
density also implies a high turnover rate of mosquito generations
that result in a constant availability of female adult mosquitoes
for disease transmission. This observations justify the treatment of 
a Dengue epidemic outbreak as if it were the epidemic outbreak in a directly 
transmitted disease. Of course, this hypothesis breaks down in the interepidemic 
period also known as the ``endemic'' state where low prevalence, asymptomatic 
infections and vector abundance are subject to stochastic environmental effects 
as well as to the seasonal forcing induced by the rainy season on the vector 
population. Consequently, we limit our estimation to the length of time 
of the epidemic outbreak in each one of the cases analyzed.

In summary we use the susceptible, infectious, recovered (SIR) Kermack McKendrick
model as an approximation for the Dengue dynamics  (e.g., Adams, Holmes, Zhang 
et al, \cite{adams2006}) for only one epidemic period. We therefore disregard demographic
influences since births and deaths may be considered to have negligible effects 
on disease dynamics during this short time period and assume the mosquito population 
dynamics is constant and plays a role as a factor in the contact rate.

We rewrite now model (\ref{eq:sir1})-(\ref{eq:sir2}) with the standard epidemiological 
notation where $b_{0}$ is the contact rate, $1/\gamma$ is the length of the infectious 
period and $N=y_1+y_2+y_3$ is the total population density:

\begin{align}
  \label{eq:sir3}
  \dot{y}_{1} &= -b_{0}\frac{y_{2}}{N}y_{1}\\
  \dot{y}_{2} &= b_{0}\frac{y_{2}}{N}y_{1} - \gamma y_{2}\\
  \label{eq:sir4}
  \dot{y}_{3} &= \gamma y_{2}.
\end{align}

The basic reproduction number for this system is
$$R_0=\frac{b_0}{\gamma},$$
where we have rescaled the total population to be $N=1$. For comparison the simplest 
vector transmitted disease based on the Ross-Macdonald model renders a reproductive number 
of the form
$$R_v=\frac{abm}{\delta\gamma'}$$
where $a$ and $b$ are the mosquito and human biting rates, $m$ is the ratio of female mosquitoes 
to human hosts and $1/\delta$ and $1/\gamma'$ are the infectiousness periods for mosquito and human
respectively.

It is obvious that the infection rate in the SIR model, $1/\gamma$, includes the total time available 
per capita for infection in both mosquito and human hosts; also the contact rate $b_0$ in the 
Kermack-McKendrick model aggregates the combined effect of the product $ab$ in $R_v$. Therefore 
the estimates of $R_0$ that we are about to obtain by applying our methods, will not discriminate 
the different components of this threshold parameter. Our methods, however, can be applied directly 
to more complicated models that explicitly incorporate the relevant parameters for the human and 
vector populations if necessary. 

\begin{figure}
\begin{center}
\includegraphics[height=6cm, width=7cm]{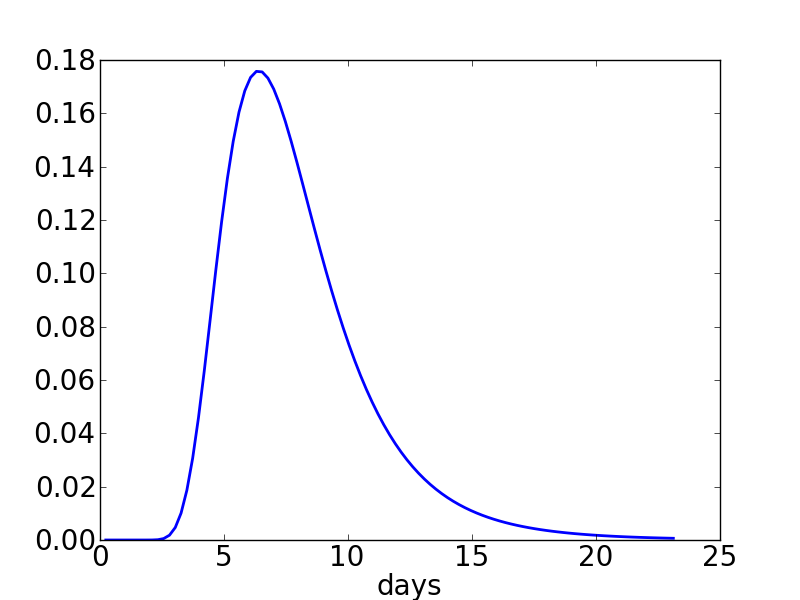}
\caption{\label{fig.DenguePriors} Inverse Gamma prior distribution for $\rho_1 = 1/\gamma = 1/b_1$ for our Dengue outbreaks examples.}
\end{center}
\end{figure}

\subsection{Acapulco 2005 and Cuernavaca 2008 outbreaks}\label{sec.exa_all}

Here we apply our methods to the two data sets described in  section (\ref{sec.proxy_data}).  
For completeness we provide now a brief description of the basic life-cycle facts of the Dengue virus.  
The endemic-epidemic cycle of Dengue has the following characteristics. After an infected human host is 
bitten the viral incubation period within the mosquito lasts about 7-14 days after which the mosquito 
becomes infectious for its entire lifespan which is at most 25 days. In the human host the incubation 
period last form 3-14 days (with an average of 7). The infectious period lasts for about 2-10 days after 
outset of symptoms. Values on biting rates are not available and these are the main components that are 
estimated or circumvented when dealing with the basic reproduction number (see Chowell, Diaz-Due\~nas, 
Miller et al, \cite{chowell2007} and Hsieh and Chen, \cite{HsiehChen2009}).  Since we do not have information for $\rho = 1/b_0$ we 
use a uniform prior over a very long range (not shown). The prior information for the infectious period 
$\rho_1 =1/b_1$ is shown in Figure~\ref{fig.DenguePriors}. Recall that $\rho_1 = 1/\gamma$ is aggregating 
the infectious period of both the human host and vector discounted by the time allocated to viral incubation. 
To see this consider the following rough approximation: take the average human infectious period of 7, and 
the mosquito infectious period of about 10 days. The total time available for transmission is 
$7 \times 10=70$;  the proportion of the mosquito and human lifespan available to infection given the 
extrinsic and intrinsic incubation periods is the weighting factor that should be incorporated into 
$1/\gamma$. Suppose an extreme scenario where all the mosquito lifespan is infectious and that the human 
intrinsic incubation period is 10 days; then roughly the proportion of time available for closing a 
transmission cycle human-vector-human is $1 \times 1/10=0.1$ which when multiplied by 70 gives the mode 
assigned to the prior for $\rho_1 = 1/\gamma$ shown in Figure~ \ref{fig.DenguePriors}.  For this inverse 
gamma distribution for $\rho_1 = 1/\gamma = 1/b_1$ the shape parameter is set to 9 and the scale parameter 
is 8 (9-1). Its mean is therefore 8 (days; these are in fact equal to $\alpha_1$ and $\beta_1$,
given the shape and \textit{scale} parametrization used for the Gamma prior for $b_1 = \gamma$ in 
(\ref{eqn.priors})).

\begin{figure}
\centering
\subfloat[]{\includegraphics[height=4cm, width=15cm]{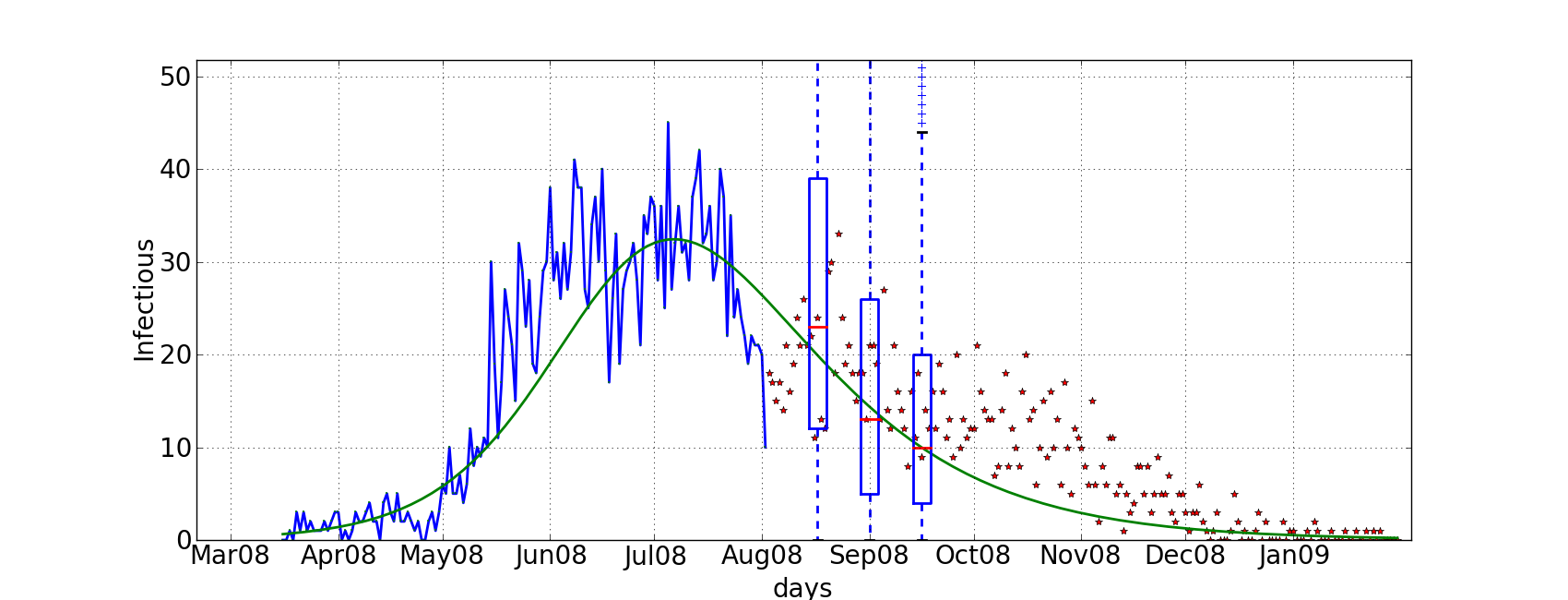}} \\
\subfloat[]{\includegraphics[height=4cm, width=5cm]{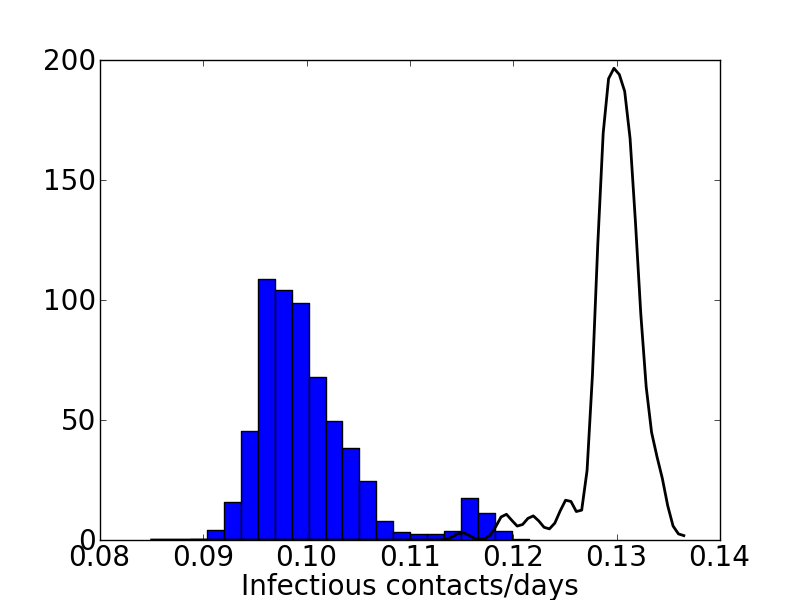}}
\subfloat[]{\includegraphics[height=4cm, width=5cm]{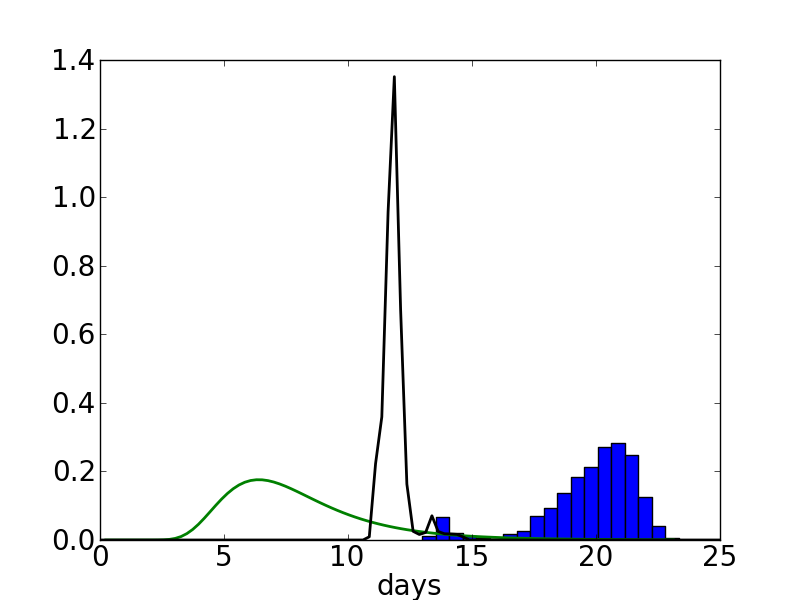}}
\subfloat[]{\includegraphics[height=4cm, width=5cm]{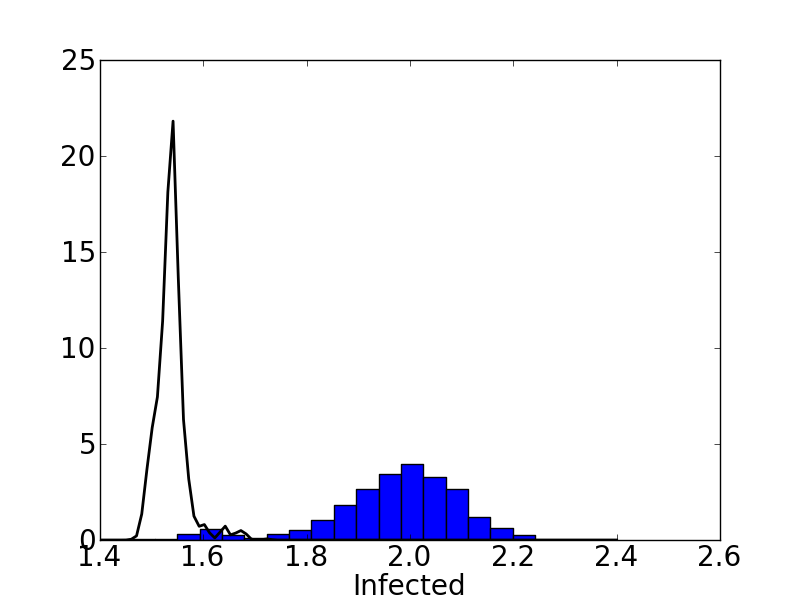}}
\caption{\label{fig.Cue08}  (a) Daily reported cases of Dengue fever in Cuernavaca, Mexico, 2008, with several
observations trimmed away at the end.  Estimated
MAP model and three future fortnights of reported cases predicted (box-plots of predictive distribution),
actual data with plotted with *.
(b) Posterior distribution for $b_0$ (c) Posterior (histogram) and prior distribution for
$\rho_1 = 1/b_1$ and (d) posterior distribution of $R_0 = b_0/\gamma$.
The approximated posterior densities with the full data are also presented (black).
}
\end{figure}

For the Cuernavaca outbreak the MAP estimator of the reproductive number $R_0$ is 1.9 with $(1.8,2.2)$ 
as HPD 95\% probability interval, when August to December 2008 are not considered in the estimation. 
The prediction of the fate of the epidemic is rather impressive, capturing basically 
all the data variability (see Figure~\ref{fig.Cue08}).  Intentionally, we chopped off data at the end
of a descending fluctuation run, after the epidemic maximum, in early August 2008.  The descending run 
did not fool our predictions of the future evolution of the epidemic (see the predictive distribution 
Box-Plots in  Figure~\ref{fig.Cue08}(a)).  At that point, reasonable estimates for $b_0$ and $b_1$ were 
already available and good predictions could be done, taking into account all modeled sources of variability. 

\begin{figure} 
\centering
\subfloat[]{\includegraphics[height=4cm, width=15cm]{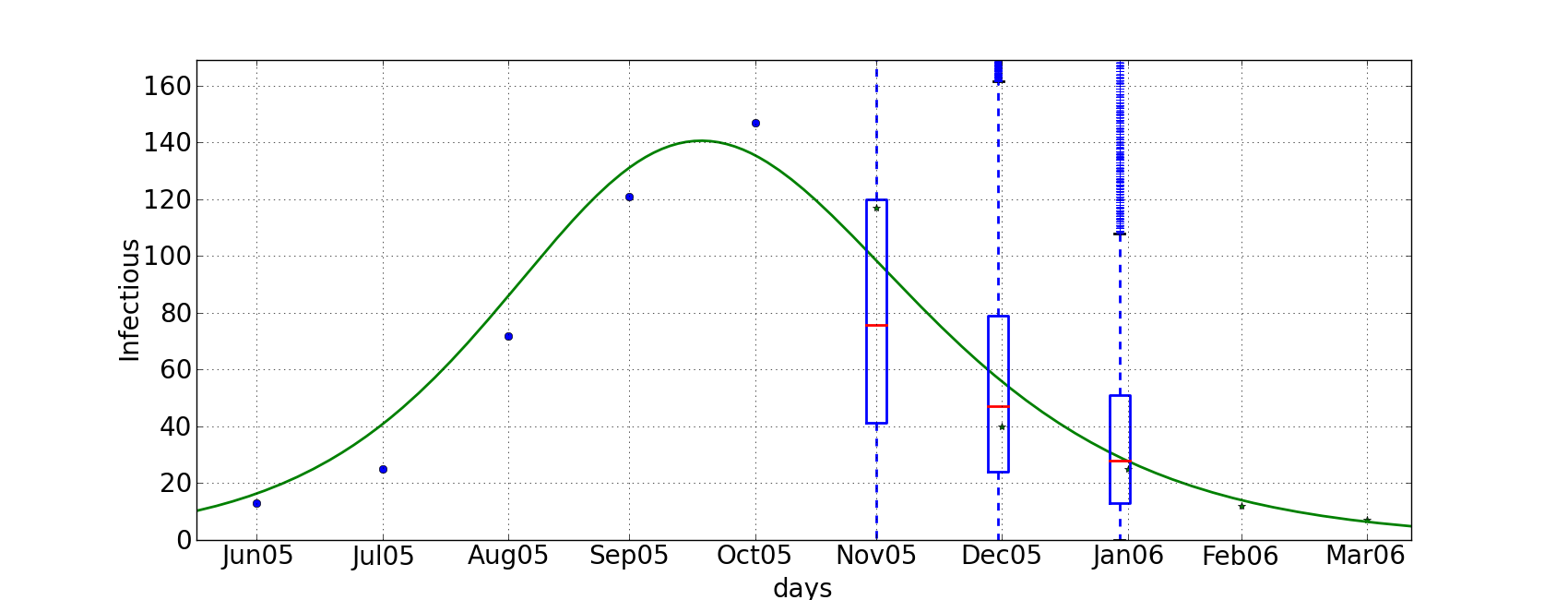}} \\
\subfloat[]{\includegraphics[height=4cm, width=5cm]{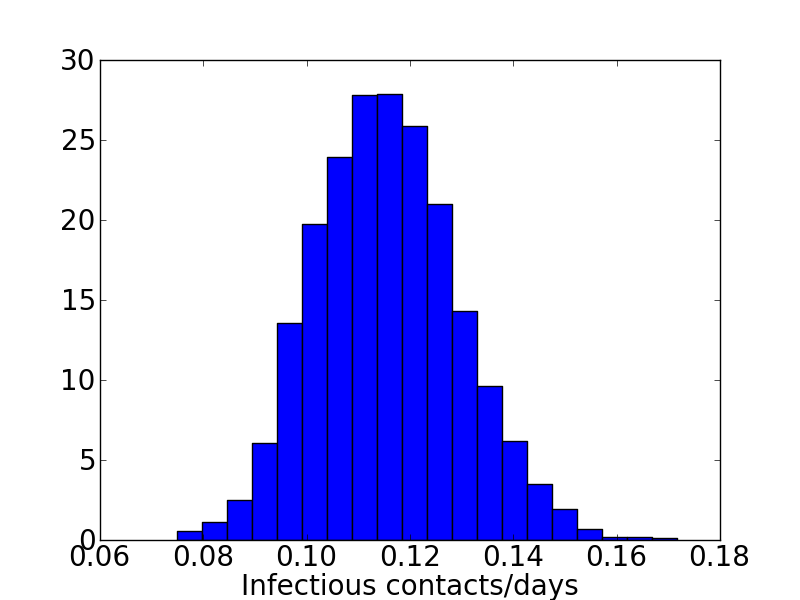}}
\subfloat[]{\includegraphics[height=4cm, width=5cm]{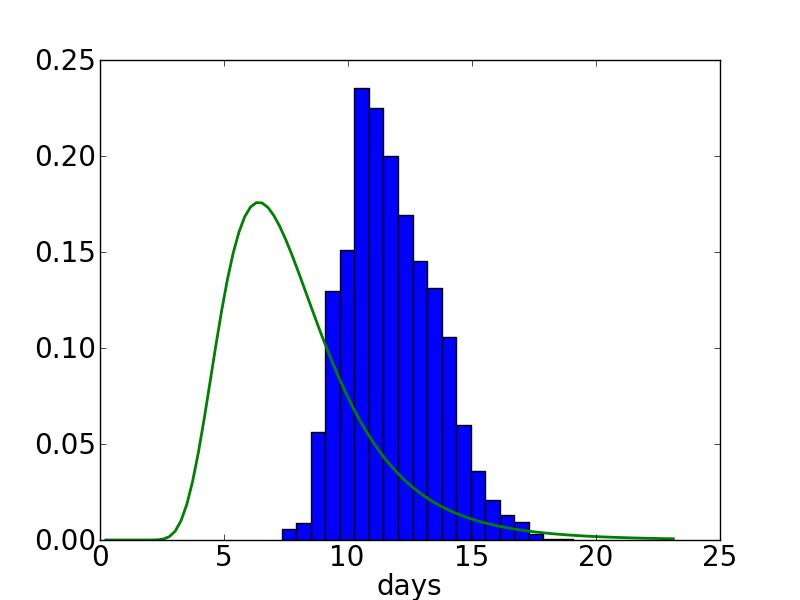}}
\subfloat[]{\includegraphics[height=4cm, width=5cm]{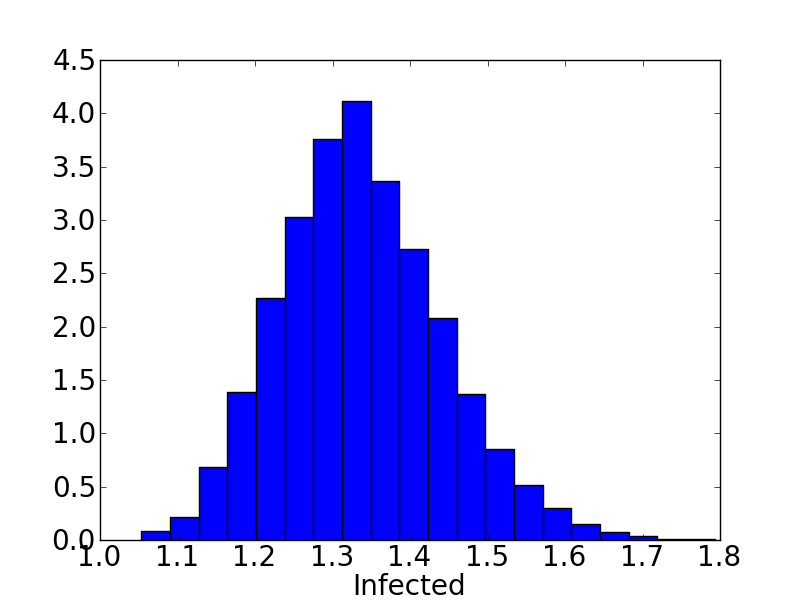}}
\caption{\label{fig.Aca05}  Reported cases of Dengue fever in Acapulco, Mexico, 2005, with 6
observations trimmed away at the end.  Estimated MAP model and three future months of reported 
cases predicted (box-plots of predictive distribution), actual data plotted with *.
(b) Posterior (histogram) and prior distribution for $\rho_0 = 1/b_0$. (c) Posterior (histogram) 
and prior distribution for $\rho_1 = 1/b_1$ and (d) posterior distribution of $R_0 = b_0/\gamma$.
}
\end{figure}

For the Acapulco outbreak the MAP estimator of the reproductive number $R_0$ is 1.3 with $(1.1, 1.6)$ as 
HPD 95\% probability interval when data starting in November 2005 are not considered in the estimation. 
The prediction of the fate of the epidemic is quite good considering the few data 
points used in the analysis (see Figure~\ref{fig.Aca05}).

The estimate for Acapulco is in the lower range of values reported in Table~\ref{tab:R0s}. In fact, 
only the lower bound for the range in Khoa et al \cite{khoa2005} is comparable in magnitude. The estimate for 
Cuernavaca is located more within the range reported by several studies. 

We remark that the scarce (monthly) data from Acapulco leads to less informative predictions, hiding 
the intrinsic stochasticity of the state variables. In contrast data from Cuernavaca are reported weekly,
See Figure~\ref{fig.Cue08}. There are two possible explanations for this trend. Either, there might be 
two waves of Dengue, one starting in early April and other developing in early August 2008. The SIR model 
is unable to capture this dynamics in the local detail (late July-early August) but it is able to capture 
the tail of the epidemic outbreak reasonably well (see Figure~\ref{fig.Cue08}a). If this hypothesis of two 
Dengue waves was true, the second wave of the epidemic corresponds may correspond to a different set of parameters, 
that increases the incidence immediately starting in August. On the other hand, the decreasing trend in
August may be explained by demographic stochasticity alone.

\section{Conclusions}\label{sec.disc}


With respect to this approach as an useful inferential and predictive framework for epidemic surveillance data,
Breban, Vardavas and Blower \cite{breban2007} present a critique to estimating $R_0$ using population level models.
These authors claim that using an individual-level model approach to estimate $R_0$ is more accurate and 
may not coincide with the $R_0$ estimated from a population level model. In the approach we present here, 
the incorporation of the CME formalism introduces demographic stochasticity, i.e., individual stochastic 
variation into the epidemic parameters $b_0$ and $b_1$ thus actually approximating individual level 
characteristics. The methods shown here do generate distributions of the basic parameters, including 
$R_0$ based on the a priori information available. To reinforce our point  the results by Lloyd-Smith, 
Schreiber, Kopp and Getz \cite{lloyd2005} show that $R_0$ has an highly skewed distribution in many diseases when 
sampled from individuals through contact-tracing. Certainly our results Figures~(\ref{fig.Cue08}) 
and~(\ref{fig.Aca05}) do show (admittedly only slightly) skewed distributions  that indicates the individual 
variability in contact and recovery rates at the individual level. This variability does not come from 
sampling errors but from an intrinsic property of stochastic individual variation constructed into the model 
via the CME. Our estimation approach uses this built-in demographic stochasticity to estimate $R_0$. We 
are using a very simple model, the SIR without vital dynamics and this may explain the relatively low 
variability reflected in our nearly symmetric distributions: the model is smoothing out individual 
differences. We are using the simplest priors that are biologically reasonable.

In general, we expect to generate highly skewed distributions for at least one 
of our parameters
when, for example, few data points are available and the uncertainty in estimating $b_1$ is such that its posterior distribution
borders on zero, the distribution of $R_0 = b_0/b_1$ will result heavily skew with a long tail (as in Figure~\ref{fig.synth2}(d)). 

Indeed, the SIR model presented here is quite simple, and specially applied to (vector induced, 
delayed, etc.) Dengue.  Nevertheless, we obtained interesting inferences and prediction power from 
limited data, in real and synthetic scenarios, and the model simplicity servers us to present and 
experiment with our inference procedure.  Our inference framework can indeed be generalized to be 
used in a more complex model (as explained in Section~\ref{sec.SIR}), were for example the vector-human 
interaction is modeled in a more comprehensive way.  We leave this for further research.

\section{Acknowledgments}

We thank the support, expert commentaries and insight into the complex dynamics of Dengue fever of Jos\'e Ramos-Casta\~neda, Ruth Aral\'{\i} Mart\'{\i}nez-Vega and Rogelio Danis-Lozano who also provided the the data on Dengue that we have used here. We also acknowledge the very fruitful discussions with the network on Biology and Ecology from CONACyT's
\textit{Red en Modelos Matem\'aticos y Computacionales}. JXVH acknowledges partial funding from the National Institute for Mathematical and Biological Synthesis through the National Science Foundation
EF-0832-858. 


\bibliographystyle{elsarticle-num}
\bibliography{inference_dengue}

\end{document}